\documentclass[twocolumn]{pasj00}

\begin{document}

\SetRunningHead{Suzaku Discovery of a Hard Component in MCG--6-30-15}{H. Noda et al.}
\title{Suzaku Discovery of a Hard Component \\
Varying Independently of the Power-Law Emission
in MCG--6-30-15}

\author{Hirofumi \textsc{Noda},\altaffilmark{1}
Kazuo \textsc{Makishima},\altaffilmark{1,2}
Yuuichi \textsc{Uehara},\altaffilmark{1}
Shin'ya \textsc{Yamada},\altaffilmark{1}
and Kazuhiro \textsc{Nakazawa}\altaffilmark{1}
}
\altaffiltext{1}{
  Department of Physics, University of Tokyo\\
  7-3-1, Hongo, Bunkyo-ku, Tokyo, 113-0033, Japan}
\altaffiltext{2}{
  Cosmic Radiation Laboratory, Institute of Physical and Chemical
  Research (RIKEN)\\
  2-1 Hirosawa, Wako-shi, Saitama, 351-0198, Japan}

\email{noda@juno.phys.s.u-tokyo.ac.jp}

\Received{$\langle$reception date$\rangle$}
\Accepted{$\langle$accepted date$\rangle$}
\Published{$\langle$publication date$\rangle$}

\maketitle

\begin{abstract}
Focusing on  hard X-ray variability,
we reanalyzed Suzaku data of Type I Seyfert galaxy MCG--6-30-15 obtained in 2006.
Intensity-sorted spectroscopy and a principal component analysis
consistently revealed a very hard component
that varies independently of the dominant power-law emission.
Although the exact nature of this hard component is not yet identified,
it can be modeled as a power-law with 
a photon index $\sim 2$ affected by a partial covering absorption,
or as a thermal Comptonization emission
with a relatively large optical depth.
When this component is included in the fitting model,
the time-averaged 2.5--55 keV spectrum of MCG--6-30-15
can be reproduced successfully by invoking a mildly broadened iron line
with its emission region  located at $\gtrsim 8$  times the gravitational radii
from the central black hole,
and a moderate reflection with a covering fraction of $ \sim 3.4 \pi$.
This result implies that the solution of a highly spinning
black hole in MCG--6-30-15,
obtained by Miniutti et al. (2007, PASJ, 59, S315) using the same Suzaku data,
is a model dependent result.

\end{abstract}

\section{Introduction}
\label{sec:intro}

Investigations of spectral and temporal characteristics
of X-ray emission from an Active Galactic Nuclei (AGN)
provides one of the most direct ways to detect
general relativistic effects  around a  black hole (BH).
Among various types of AGNs,
Seyfert galaxies are known to exhibit an X-ray spectrum
which can be decomposed mainly into a variable power-law continuum,
iron lines, and a reflection component.
Using ASCA,  \citet{Tanaka1995} detected a broad iron line feature
from  the Type I Seyfert galaxy MCG--6-30-15,
and interpreted its extreme broadening as a result of
special and general relativistic effects near the  BH
residing in this AGN.
Later, this result was strengthened by other satellites,
including XMM-Newton and Chandra,
followed by reports of the detections of similar features
from other Seyfert galaxies (for reviews, \cite{ReynoldNowak2003}; \cite{Nandra2007}).

Utilizing the broad-band coverage of Suzaku,
\citet{Miniutti2007} succeeded in reproducing
the ASCA results on MCG--6-30-15.
Specifically, they detected an apparently broad Fe-K line
using the X-ray Imaging Spectrometer (XIS; \cite{Koyama07}),
and a strong reflection hump over an energy range of 20--40 keV
using  the Hard X-ray Detector (HXD; \cite{Takahashi07}).
The large equivalent width of the  former, $\sim 320$ eV,
is consistent with the large reflection fraction,
$f_{\rm refl} \equiv \Omega / 2 \pi \sim 4$, of the latter
(with $\Omega$ being the reflector's solid angle).
A quantitative modeling of  these reprocessed components
by \citet{Miniutti2007} indicated
that they are emitted from an accretion disk
with an inner radius of $R_{\rm in} \sim 2\;R_{\rm g}$.
Here,  $R_{\rm g}$ is the gravitational radius,
and is $\sim 15$ lt s in the present case assuming a BH mass of
$3 \times 10^{6}\;M_{\odot}$ (McHardy et al. 2005).
Because the derived $R_{\rm in}$ is significantly smaller
than the radius of the last stable circular orbit ($6\;R_{\rm g}$)
around a non-spinning BH,
the BH in this Seyfert galaxy was concluded
to be rapidly spinning \citep{Miniutti2007}.

During these Suzaku observations,
the iron line flux and the reflection-component intensity were
both approximately constant on time scales of 45 ks \citep{Miniutti2007},
while the power-law emission did vary.
Although this behavior apparently contradicts the idea
that the iron line and the reflection component are
emitted from a close vicinity of the black hole,
\citet{Miniutti2007} introduced the  ``light-bending" model 
as a possible explanation to the phenomena:
when the power-law emitting region is located at
particular positions above the accretion plane,
and  gets closer to the BH,
the power-law component becomes dimmer
because more photons are swallowed into the black hole,
while the strength of the reflected component is kept rather constant.
In addition, this geometry can explain the unusually large value of $f_{\rm refl}$.

The light-bending model, however, requires the specific geometry with fine tuning,
in order to explain the varying power-law contimuum
and the non-varying reprocessed components \citep{NiedzwieckiZycki2008}.
In addition, some other works showed
that the same Suzaku spectrum of MCG--6-30-15 can be reproduced
with alternative spectral modelings,
invoking ionized absorbers and partial covering,
but using neither the broad iron line nor the strong reflection
(e.g., Miller et al. 2009, \cite{Miyakawa2009}).
Thus, the same data allow degenerate interpretations
with different physical  implications,
one requiring a rapid BH rotation whereas the others  not.

To disentangle the model degeneracy,
identifying the underlying spectral components
in a model independent manner is critically important.
For this purpose, many works, such as statistical and timing analyses,
have been performed so far
(e.g., \cite{Taylor2003,Vaughan2004, L_Miller09}).
However, due to lack of sensitivity in the energy band in $\gtrsim 15$ keV,
few of them focused on the hard X-ray variability.
In the present paper,
we hence studied the variability of MCG--6-30-15 in the  hard X-ray band,
where the  putative large reflection is dominant.
As a result, we have successfully identified a very hard component
that may be distinct from the usually assumed reflection.
Taking this component into account,
we have succeeded in reproducing the time-averaged 2.5--55 keV spectrum
without invoking  the extremely broad iron line or the strong reflection.

\section{Observation and Data Reduction}
\label{sec:obs}

The present work uses the same Suzaku data of
MCG--6-30-15 as Miniutti et al. (2007) did.
They were obtained with the XIS and the HXD on three occasions,
9--14, 23--26, and 27--30 of January 2006.
After standard data screening process of Suzaku,
the exposures achieved with the XIS  are
143 ks, 98 ks, and 97 ks in the three observations,
while those with HXD-PIN are 126 ks, 82 ks, and 90 ks, respectively.
We analyze on-source and background data
processed via version 2 pipeline processing,
together with the corresponding software and calibration data.

On-source events of each XIS camera were extracted
from a circular region of $6'$ radius centered on the source,
while background XIS events were taken from a surrounding annular region
of the same camere with the inner and outer radii of $6'$ and $10'$, respectively.
We added events from XIS0, XIS2, and XIS3,
because these three front-illuminated (FI) CCDs
have almost identical responses.
The response matrices and ancillary response files were created
utilizing xisrmfgen and xissimarfgen (Ishisaki et al. 2007), respectively.
The data from XIS1 (back-illuminated CCD) are not utilized
in the present work.

In a similar way, events from HXD-PIN were accumulated.
Non X-ray background (NXB), included in the on-source data,
is estimated by analyzing the set of fake events
which are created by a standard NXB model (Fukazawa et al. 2009).
By analyzing the on-source and the NXB event sets in the same way,
we can subtract the NXB contribution from the HXD-PIN data.
The events remaining after the NXB subtraction still include
contributions from the cosmic X-ray background (CXB; Bold et al. 1987),
which must be subtracted as well.
The CXB contribution was estimated using the HXD-PIN response to diffuse sources, 
assuming the spectral CXB surface brightness model
determined by HEAO 1 (Gruber et al. 1999):
$9.0 \times 10 ^{-9} (E/3~\mathrm{keV})^{-0.29} \exp (-E/40~\mathrm{keV})$
erg cm$^{-2}$ s$^{-1}$ str $^{-1}$ keV$^{-1}$,
where $E$ is the photon energy.
The estimated CXB count rate is 5\% of the NXB signals.

\section{Timing Analysis}
\label{sec:timing}

\subsection{Light curves}

\begin{figure*}[t]
 \begin{center}
  \FigureFile (200mm,200mm)
    {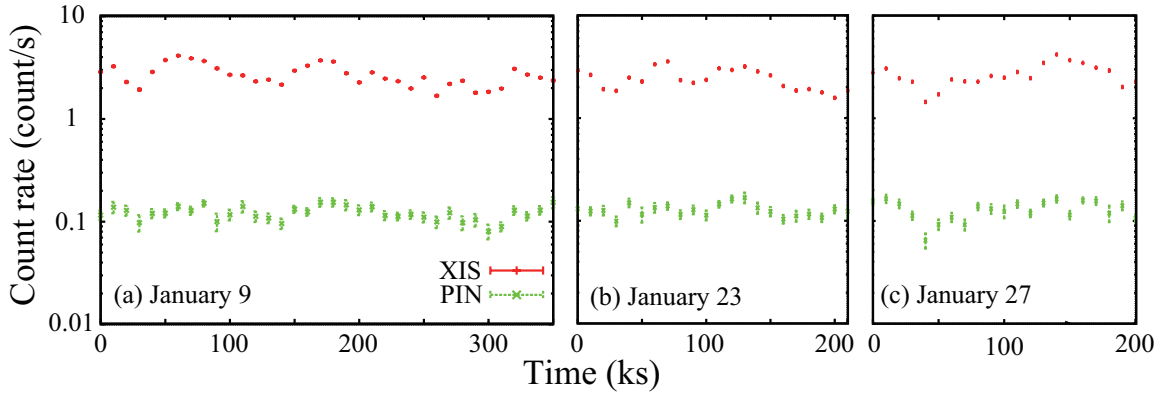}
 \end{center}
 \setlength{\abovecaptionskip}{-30mm}
\setlength{\belowcaptionskip}{0mm}
 \caption
{Background-subtracted and dead-time corrected
light curves of MCG--6-30-15,
measured with the XIS (3--10 keV; red)
and HXD-PIN (15--45 keV; green),
shown with a binning of 10 ks.
Panels (a), (b) and (c) represent
the data on January 9, January 23,
and January 27, respectively.
Error bars represent statistical $1\sigma$ ranges.
}
\label{fig:lc}
\end{figure*}

Figure~\ref{fig:lc} shows light curves of MCG-6-30-15,
obtained on the three occasions with the XIS and HXD-PIN.
All of them are shown after subtracting the backgrounds
(both NXB and CXB) as described in section~\ref{sec:obs},
and corrected for dead times.
In addition to clear variations by a factor of 2 in the XIS light curves,
those of HXD-PIN also vary significantly
with a typical amplitude of $\sim \pm 30\%$ ($\sim \pm 0.03$ cnt s$^{-1}$).
Since the HXD-PIN NXB is modeled within
a systematic uncertainty of $1-2\%$ (Fukazawa et al. 2009),
or $0.003-0.006$ cnt s$^{-1}$ in the present case,
the HXD-PIN variations are real rather than instrumental.
Although the hard X-ray variations generally follow
those in the soft X-rays,
the two bands sometimes appear to be varying in
rather uncorrelated manner.
This issue is discussed in the next subsection.

\subsection{Count-Count Plots}
\label{subsec:CCP}

\begin{figure*}[t]
 \begin{center}
   \FigureFile(180mm,180mm)
    {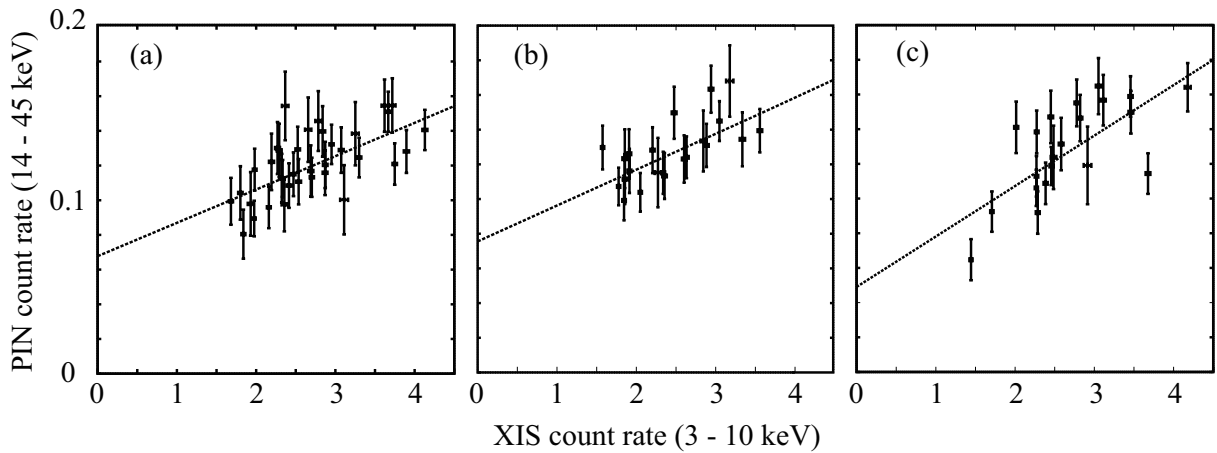}
 \end{center}
  \setlength{\abovecaptionskip}{-35mm}
\setlength{\belowcaptionskip}{0mm}
 \caption
{Count-Count Plots of MCG--6-30-15 obtained on
(a) 2006 January 9, (b) 2006 January 23, and (c) 2006 January  27.
Abscissa gives NXB-inclusive XIS (3--10 keV) count rates,
while ordinate gives those of  HXD-PIN (15--45 keV)
after the NXB subtraction, both with a binning of 10 ks. }
\label{fig:ccp}
\end{figure*}

In order to examine to what extent the XIS and HXD intensities are correlated,
we made  scatter plots between the 3--10 keV XIS count rates
and those of HXD-PIN in 15--45 keV, and show them in figure~\ref{fig:ccp}.
We hereafter call them Count-Count Plots (CCPs).
There, the time bin size is chosen to be 10 ks,
which is a minimum integration time needed to achieve sufficient photon statistics.
These CCPs employ the XIS count rates with background subtraction.
The HXD data are used in figure~\ref{fig:ccp} with the NXB also subtracted,
but with the CXB inclusive,
because the CXB should be constant with time,
and only gives a positive offset by $\sim 0.02$ cnt s$^{-1}$.
The errors associated with  the HXD data points are
quadratic sum of the statistical errors
and the systematic NXB error by $\sim 1.4\%$ (Fukazawa et al. 2009).
In all CCPs, the XIS and HXD count rates both  vary by a factor of $\sim 2$,
generally in positive correlations.

As shown in figure~\ref{fig:ccp},
we fitted the data points in each CCP with a straight line.
The fits in the first and the second observations were both acceptable,
with $\chi^2/$d.o.f. = $33.38/34$ and $\chi^2/$d.o.f. = $18.69/20$, respectively.
Therefore, the soft and hard intensities in these data sets
are consistent with being fully correlated,
without evidence of any hard-band variations
that is independent of those in the soft X-ray band.
In other words,
the variability in these data sets is ``one-dimentional".
The obtained linear fits are described as
\begin{eqnarray}
&y= (0.019 \pm0.003)x+ (0.068 \pm 0.010)~~&(\mathrm{1st~obs.}), \\
&y= (0.021 \pm0.005)x+ (0.076 \pm 0.012)~~&(\mathrm{2nd~obs.}),
\label{eq:ccpfit12}
\end{eqnarray}
where $x$ and $y$ denotes the XIS and HXD-PIN counts, respectively.
Thus, the hard counts have a positive offset
that is larger than is predicted by the CXB contribution ($\sim 0.016$ cnt s$^{-1}$).

Similarly, the best-fit linear relation to the third observation was obtained as
\begin{equation}
y= 0.029x + 0.049~~. \\
\label{eq:ccpfit3}
\end{equation}
However,  the fit  is unacceptable with $\chi^2/$d.o.f. = $45.12/19$.
[As a result, the errors in equation (3) cannot be evaluated appropriately.]
Thus, the hard X-ray variation in this data set is
partially independent of those in the soft band,
as suggested by figure~\ref{fig:lc}.
In other words, the overall variation  involves
at least two independent parameters.
This inference is reinforced in  Appendix 1 using principal component analysis.
This independent hard X-ray variability has not been reported in previous studies of this AGN.

\subsection{Examination of the NXB reproducibility}
\begin{figure*}[t]
 \begin{center}
   \FigureFile(150mm,150mm)
    {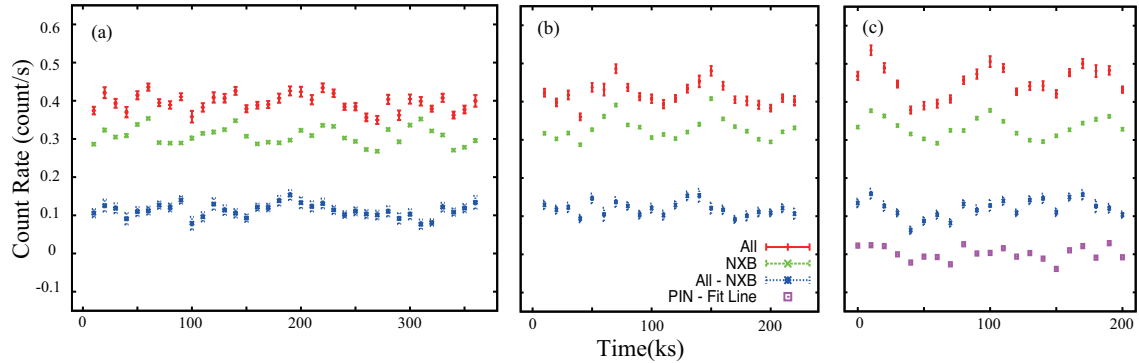}
 \end{center}
 \caption
{Detailed 15--45 keV HXD-PIN light curves of the three observations with a 10 ks binning.
The raw count rate is in red, the modeled NXB is in green, 
and the NXB-subtracted signal is in blue.
Purple in panel (c) shows the hard X-ray deviation in  figure 2c from the correlation line.}
\label{fig:lcs}
\end{figure*}
\begin{figure*}[t]
 \begin{center}
   \FigureFile(100mm,100mm)
    {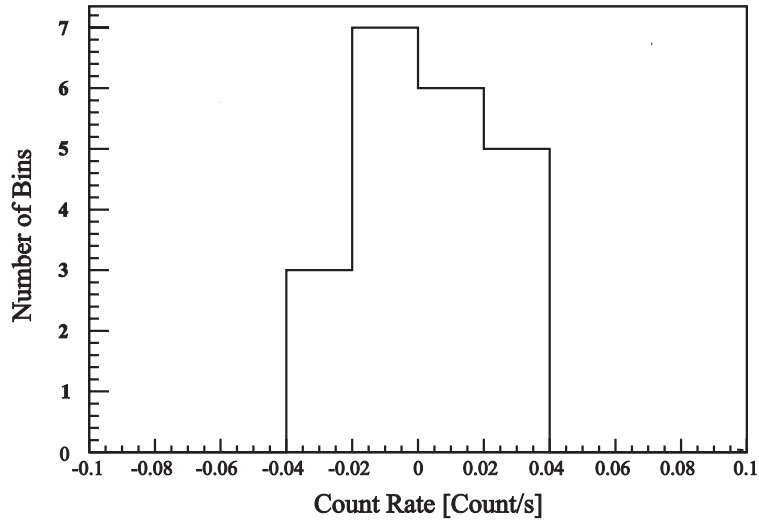}
 \end{center}
 \caption
{A histogram of the hard X-ray deviation from the best-fit line in the CCP of the 3rd observation (figure 2c). }
\label{fig:hist}
\end{figure*}

\begin{figure*}[t]
 \begin{center}
   \FigureFile(170mm,170mm)
    {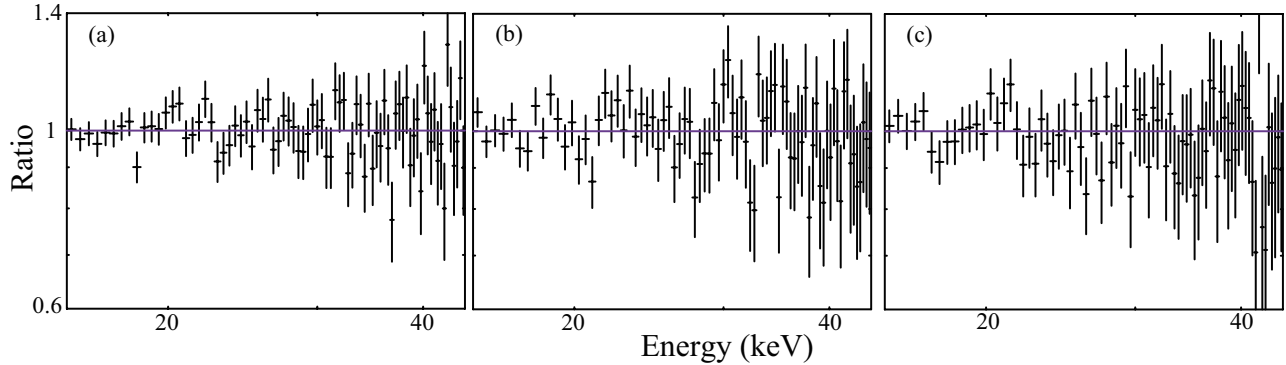}
 \end{center}
 \caption
{The ratios between the HXD-PIN spectrum accumulated over Earth occultation periods 
and the modeled NXB spectrum on
(a) 2006 January 9, (b) 2006 January 23, and (c) 2006 January  27.}
\label{fig:ratio}
\end{figure*}

\begin{figure*}[t]
 \begin{center}
   \FigureFile(100mm,100mm)
    {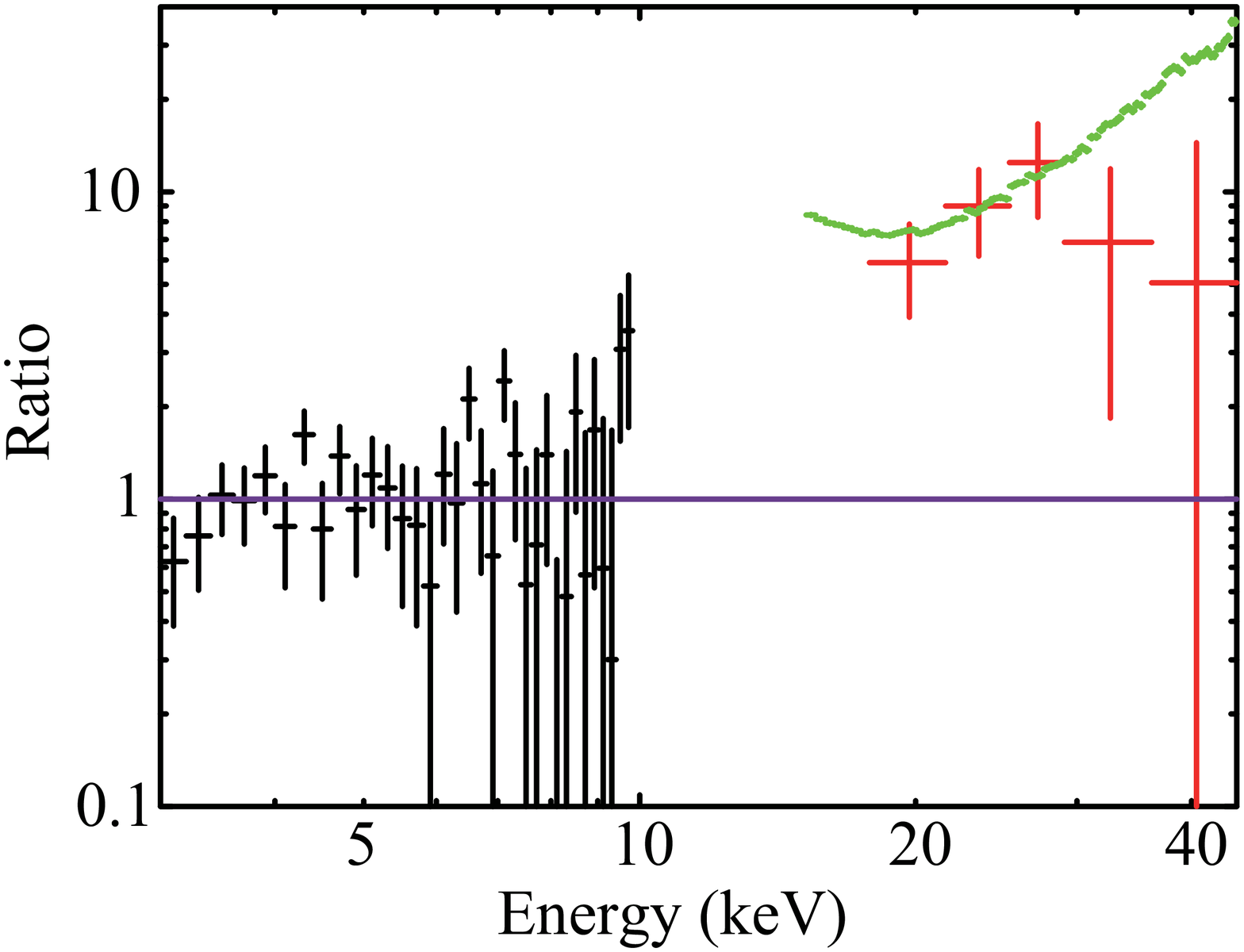}
 \end{center}
 \caption
{The difference spectrum from the 3rd observation,
divided by the prediction of a  $\Gamma = 2.0$ power-law model.
A green curve shows $5\%$ of the HXD-PIN NXB.}
\label{fig:diffspec}
\end{figure*}

There would be  an immediate suspect
that the gtwo-dimensionalh data scatter in figure 2c 
simply reflects variations in the HXD-PIN background
that have not been accurately subtracted out by the NXB model.
To investigate this issue, 
we compare in figure 3  light curves (with 10 ks binning) 
of four relevant quantities (see caption for the explanation).
If the effect seen in figure 2c were due to residual NXB variations,
the purple light curve in figure 3c should show 
some (anti-) correlations to the NXB model (green),
because  the NXB model reproducibility  is limited  
mainly by the accuracy to account for orbit-related 
semi-periodic NXB changes (Fukazawa et al. 2009).
However, we do not find such correlations in figure 3c.

Figure 4 shows an occurrence distribution 
of the purple light curve in figure 3c.
(This is identical to the distribution of data deviations
in figure 2c from the correlation line).
This histogram exhibits an rms scatter by $0.017\pm0.003$cnt s$^{\rm -1}$
which corresponds to 5.2\% of the average NXB rate,
around the mean of $0.0018$cnt s$^{\rm -1}$.
This is significantly lager than the typical rms scatter
of the NXB modeling residuals at the 10 ks binning, 
2.3\%, as reported by Fukazawa et al. (2009).
Therefore, the  21 data points,
constituting the HXD-PIN count rate in figure 3c purple,
are found to exhibit a standard deviation 
which is 2.3 times larger than that of the population
formed by the residual NXB rate at 10 ks binning.
According to $F$-tests,
we find a  chance probability of less than 1\%
for the former sample to be derived  from the latter population.
We hence conclude that  the effect observed in figure 3c
is difficult to explain in terms of residual NXB fluctuations.

There would still remain a concern
that the NXB behavior was not normal during the 3rd observation.
To examine this possibility, 
we accumulated the HXD-PIN spectrum over Earth 
occultation periods in the three observations,
with an exposure of 94 ks, 52 ks, and 42 ks,
and normalized them to those predicted by the NXB model.
As shown in figure 5,
the three spectral ratios behave in very similar ways,
all close to unity, without any indication that the 3rd observation was unusual.
In particular, the 3rd ratio (as well as the other two) is very flat towards the highest and lowest energy limits,
where any anomalous NXB behavior would appear preferentially. 

From these examinations, 
we conclude that the secondary hard component in the 3rd data set is not an 
artifact caused by uncertainties in  the NXB reproducibility.
In addition, this component is not due either 
to slow spectral variations associated,
e.g., with long-term changes in the accretion geometry,
because the purple light curve in figure 3c varies 
on a typical time scale of several tens kilo seconds.

\subsection{A Variable Hard Component}
\label{subsec:diffspec}

\begin{figure*}[t]
 \begin{center}
   \FigureFile(130mm,130mm)
    {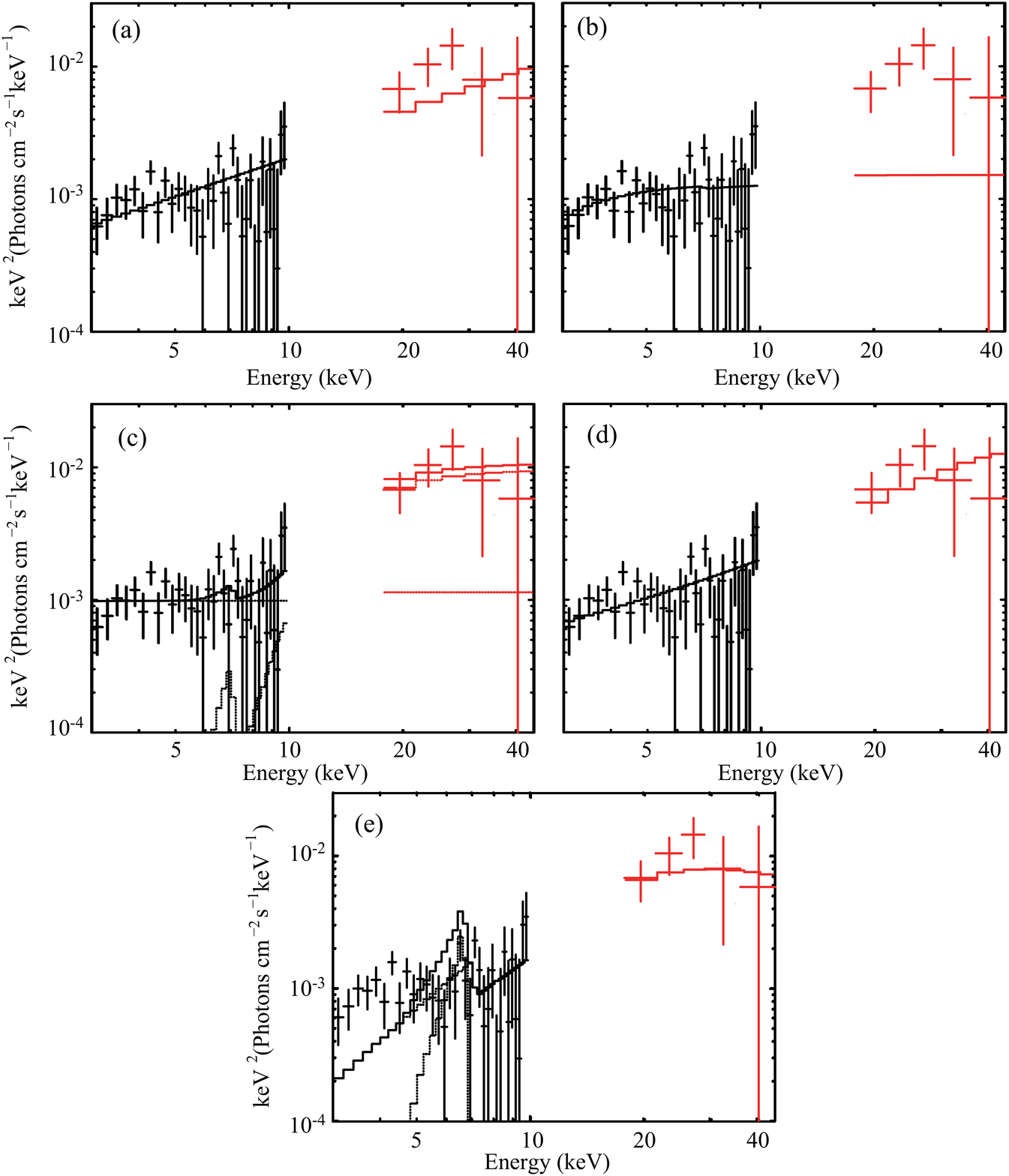}
 \end{center}
\setlength{\abovecaptionskip}{-30mm}
\setlength{\belowcaptionskip}{0mm}
 \caption
{The difference spectrum in the $\nu F_{\nu}$ form from the 3rd observation,
fitted with
(a) an unabsorbed power-law with a free slope,
(b) a power-law with $\Gamma = 2.0$ absorbed by an ionized medium,
(c) a partially absorbed power-law with $\Gamma = 2.0$,
(d) a thermal Comptonization spectrum with an electron temperature of 13.4 keV,
seed photon temperature of 0.1 keV, and an optical depth of 12.0,
and (e) reflection plus a broad Fe-K line.}
\label{fig:diffspec_fits}
\end{figure*}

Having confirmed that the secondary hard X-ray variation in the 3rd
observation is real rather than instrumental, let us examine how the 
wide-band spectrum changes as the hard X-rays vary independently 
of the soft X-rays.
To see this, we defined ``High" and ``Low" phases,
wherein the data points are above and below
the straight line in the CCP, respectively.
We then subtracted the spectrum accumulated
over the ``Low" phase, from that summed over  the ``High" phase.
As a result of this subtraction,
positive signals have remained
in both the XIS and PIN bands.
Figure \ref{fig:diffspec} shows  ``difference spectrum" obtained in this way,
normalized to the prediction of a power-law model of photon index $\Gamma = 2.0$.
The ratio, though approximately constant in the XIS band,
increases by an order of magnitude in the HXD-PIN band.
We interpret that  the data points in the CCP
of the 3rd observation (figure 2c) exhibit the vertical scatter,
because this hard spectral component varied,
on time scales of 10--50 ks,
independently of the primary power-law component.

In figure~\ref{fig:diffspec}, we also show the HXD-PIN NXB spectrum,
reduced in normalization to 5\% of its average.
Although it is thus similar in shape to the difference spectrum,
the latter does not increase to lower energies
where the NXB uncertainty is largest (Fukazawa et al. 2009).
Furthermore, the NXB spectrum rises toward higher energies,
while the other does not.
Therefore, it is hard to explain the difference spectrum
in terms of  insufficient NXB subtraction. 
This reinforces our conclusion made in subsection 3.3.

To quantify the difference spectrum,
we fitted  it with several models.
Figure~\ref{fig:diffspec_fits}a shows the simplest case of
fitting it with an unabsorbed power-law model.
The fit was in fact  acceptable with $\chi^{2}/$d.o.f. = $44.47/41$,
and gave  $\Gamma = 1.03^{+0.49}_{-0.23}$.
Thus, the difference spectrum indeed has a much harder average slope
than the primary $\Gamma \sim 2$ power-law
which dominates the average spectrum of MCG--6-30-15.
However, this $\Gamma = 1$ fit is not necessarily successful 
if the HXD data only are considered.
This urges us to look for better modeling.
This hard slope may be a result of absorption
of a certain fraction of the primary  component,
and changes of this fraction may be responsible
for the independent hard X-ray variation.
To examine this possibility,
we refitted the difference spectrum with a $\Gamma = 2.0$ (fixed) power-law,
absorbed by an ionized medium with a free column density $N_{\rm H}$
and free ionization parameter $\xi$.
The chemical abundance of the ionized absorber was fixed at 1.0 solar.
This modeling gave a column density of $10^{23}$ cm$^{-2}$ 
and $\log \xi = 0.70$ as the best estimates,
but the fit  was unsuccessful  with $\chi^2/$d.o.f. = $50.24/41$,
because the model  fell much short of the HXD-PIN data
as presented in figure~\ref{fig:diffspec_fits}b.
Therefore, a simple ionized absorber interpretation is not  appropriate:
when assuming a $\Gamma = 2.0$ power-law as the incident spectrum, 
it cannot explain the slope hardening in the 10--20 keV interval. 

A simple and promising way to improve the above two unsuccessful modelings,
and to create an apparently very hard continuum with low-energy flattening,
is to assume  ``partial covering" of the primary power-law. 
When a thick material absorbs a certain fraction
of the emission while the rest reaches us without absorption, a spectrum like
figure~\ref{fig:diffspec} will emerge. 
Indeed, the presence of such a partially-covering absorber (with possible ionization)
is supported by various observations (e.g., Miller et al. 2009, Miyakawa et al. 2009).
We have fitted the different spectrum with a partial covering model,
 \texttt{wabs * power + wabs * power}.
The photon indices of the two power laws were both fixed at 2.0, 
and one \texttt{wabs} was fixed at the Galactic line-of-sight value toward this AGN, 
$4.0 \times 10^{20}$cm$^{-2}$.
As a result, the fit was very successful with $\chi^{2}$/d.o.f = $33.35/41$; 
the thicker column density became $N_{\rm H} = 2.5^{+7.3}_{-0.7} \times 10^{24}$ cm$^{-2}$,
and the ratio between the strongly absorbed power-law to the other one was  $0.89 \pm 0.06$. 
As shown in figure 7c, the dominant strongly-absorbed signals explain the HXD data, while the minor ``leak through" signals mainly account for the XIS data.

This variable hard component  must originate in regions near the central BH,
because its variational time scale of $\sim 10$ ks
translates to a distance of $\sim 700\;R_{\rm g}$.
Then, thermal Comptnization is one of alternative production
processes of such a hard continuum in such regions,
as evidenced by the most recent Suzaku results on Cyg X-1
which revealed the Compton corona to have multiple optical depths (Makishima et al. 2008).
We therefore fitted the difference spectrum with
a thermal Comptonization model, \texttt{comptt} in Xspec12,
with the seed photon temperature  fixed at 0.1 keV
and the coronal geometry assumed to be disk.
As presented in figure~\ref{fig:diffspec_fits}d,
the fit has been successful with $\chi^2/$d.o.f. = $40.69/41$,
yielding an optical depth of $\tau = 5.3^{+1.7}_{-1.8}$
and an Comptonizing electron temperature of
$T_{\rm{e}} = 16.0_{-7.1} ^{+68.4}$ keV.
Therefore, this variable hard component allows an alternative interpretation
as a result of thermal Comptonization in a relatively thick and cool corona.
Even if this component is absorbed, e.g., 
by a neutral column of $N_{\rm H}=1.0 \times 10^{22}$ cm$^{-2}$,
the fit parameters remain unchanged within the errors.

The difference spectrum,
though reproduced with a partial covering model or \texttt{comptt} ,
could  still be a part of reflection,
considering its  resemblance in shape to the ordinary reflection component.
To examine this possibility,
we fitted the difference spectrum alternatively with an iron line and a reflection,
modeled by \texttt{laor} (Loar 1991)
and \texttt{pexrav}  \citep{MagdziarzZkziarski1995}, respectively.
The  \texttt{laor} component was assumed to have typical parameters;
the center energy of 6.4 keV, the emissivity index of 3.0, $R_{\rm in} = 10\;R_{\rm g}$,
and an inclination of 30$\degree$.
Its equivalent width was set at 1500 eV with respect to \texttt{pexrav},
which we would expect when a power-law of $\Gamma = 2.0$ 
illuminates a reprocessing matter of solar abundance 
with a solid angle of $2\pi$ \citep{GeorgeFabian1991}.
However, as shown in  figure~\ref{fig:diffspec_fits}d,
the fit was unsuccessful ($\chi^2/$d.o.f. = $120.14/41$),
mainly due to the lack of Fe-K edge and line features in the data,
and a model deficit in the 3--5 keV range.
Therefore, the difference spectrum cannot be interpreted
in terms of a reprocessed (reflection and iron line) component.
Changing the \texttt{laor} parameters did not help very much.

\section{Reanalysis of time-averaged spectra \\of MCG--6-30-15}
Through the CCP and difference-spectrum techniques,
we have shown that the hard X-ray variation of MCG--6-30-15
during the 3rd Suzaku observation is driven not only by intensity variations of the primary 
(unabsorbed) power-law continuum, but also by those of a secondary component with a hard spectrum.
Then, we expect to find the same secondary component in the time-averaged spectra. 
In figure 2, the CCPs in the 1st and 2nd observations are
consistent with exhibiting one-dimensional distributions, 
unlike the 3rd observation.
This may be understood by presuming 
that  the secondary hard  spectral component was present on all three occasions,
but its intensity varied significantly only in the 3rd observation.
With this in mind, we analyze in this section the 2.5--55 keV spectrum of MCG--6-30-15, 
summed over the three observation to increase statistics. 

\subsection{Fitting including the variable hard component}

\begin{figure*}[t]
  \begin{center}
    \FigureFile(150mm,150mm)
     {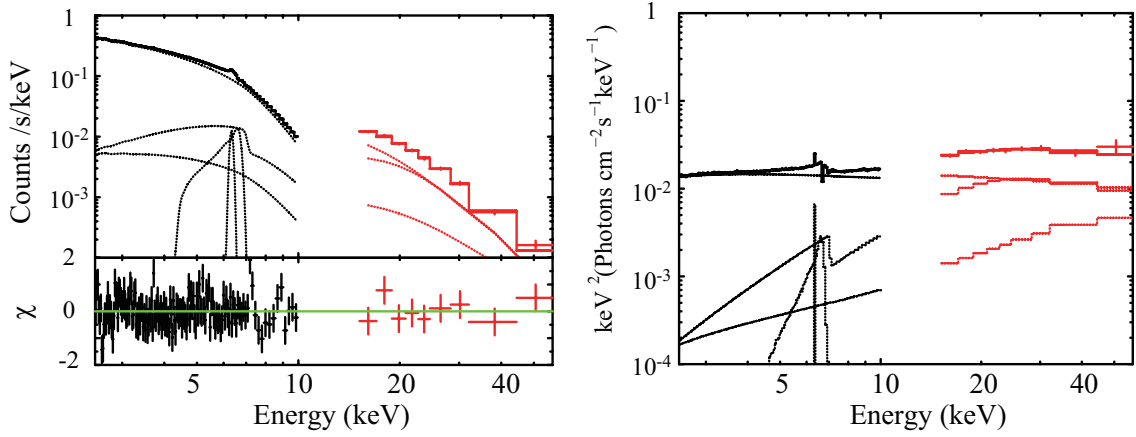}
  \end{center}
  \caption
 {Results of fitting the time-averaged spectra of MCG--6-30-15 with  
 \texttt{wabs * (cutoffpl + pexrav + comptt + laor + three narrow gaussians)} model.
 The left panel shows the data and the best-fit model prediction, while the right panel their incident $\nu F_{\nu}$ forms.}
 \label{Test}
\end{figure*}



\begin{table*}[t]
 \caption{The best fit parameters of MCG --6-30-15 with the model
 \texttt{wabs * (cutoffpl + pexrav + comptt + laor + three narrow gaussians) (3rd column), 
 or by further convolving pexrav with kdblur (4th column).} }
 \label{all_tbl}
 \begin{center}
  \begin{tabular}{llrr}
   \hline\hline
   Component & Parameter & W/o kdblur & With kdblur \\

   \hline
  wabs    & $N_{\rm H}^{*}$%
                         & $0.78^{+0.04}_{-0.05}$
                         & $0.95^{+0.05}_{-0.05}$\\[1.5ex]
   
   powerlaw     & $\Gamma$%
                       & $2.13^{+0.06}_{-0.09}$
                       & $2.24^{+0.11}_{-0.12}$\\

                    & $N_\mathrm{PL}^{\dagger}$%
                       & $1.87^{+0.01}_{-0.03}$
                       & $2.12^{+0.02}_{-0.02}$ \\[1.5ex]
                       
     reflection  & $f_\mathrm{ref}$%
                         & $1.2^{+0.2}_{-0.1}$
                         & $1.7^{+0.3}_{-0.2} $\\
                         & $A_\mathrm{Fe}$%
                         & $1.21^{+0.28}_{-0.2}$
                         & $0.98^{+0.08}_{-0.13}$\\[1.5ex]
                          
   \texttt{comptt}  &$T_\mathrm{e}$(keV)
                             &$16.0$(fix)
                             &$16.0$(fix)\\
                             &$\tau$
                             &$5.3$(fix)
                             &$5.3$(fix)\\
                         &$N_\mathrm{C}^{\#}$
                         &$4.07^{+5.08}_{-4.07}$
                         &$7.42^{+3.64}_{-4.10}$ \\  [1.5ex]

 Fe~I~K$\alpha$   & $E_\mathrm{c}^{\ddagger}$ %
                        & $6.38^{+0.17}_{-0.17}$
                        & $6.35^{+0.08}_{-0.13}$\\

                    & $R_\mathrm{in} (R_\mathrm{g})$%
                        & $8.24^{+2.83}_{-2.27}$
                        & $9.47^{+1.65}_{-1.93}$ \\
                                       
                    & $i$ (degree) %
                        & $33.2^{+12.3}_{-7.9}$
                        & $35.9^{+4.1}_{-1.7}$ \\    
                        
                   & $N_\mathrm{Fe}^{\S}$ %
                        & $5.50^{+0.92}_{-0.92}$
                        & $6.86^{+1.09}_{-0.89}$ \\                    

                    & $EW$ (eV) %
                        & $145^{+93}_{-74}$
                        & $196^{+65}_{-57}$\\[1.5ex]

   $\chi^{2}$/d.o.f. &   & 172.44/157  & 150.84/157\\
      \hline\hline
      
  \end{tabular}
 \end{center}
   
  	{\small
         \footnotemark[$*$] Equivalent hydrogen column density in  $10^{22}$ cm$^{-2}$ \\
         \footnotemark[$\dagger$] The power-law normalization at 1 keV, in units of $10^{-2}$~photons~keV$^{-1}$~cm$^{-2}$~s$^{-1}$~at 1 keV \\
         \footnotemark[$\ddagger$] Center energy in keV\\
         \footnotemark[$\S$] The broad iron line normalization in units of 
         $10^{-5}$~photons~keV$^{-1}$~cm$^{-2}$~s$^{-1}$\\
          \footnotemark[$\#$] The thermal Comptonization component normalization in units of 
         $10^{-5}$~photons~keV$^{-1}$~cm$^{-2}$~s$^{-1}$. }

\end{table*}


To look for possible contributions of the variable hard component in the persistent X-ray emission, 
we fitted the summed time-averaged spectra of XIS and HXD-PIN of MCG--6-30-15 with a model  
\texttt{wabs * (cutoffpl + pexrav + comptt + laor + three narrow gaussians)}. 
Although the secondary component is expressed here in terms of the Comptonization scenario,
this is for simplicity only, and the results to be obtained below are essentially the same 
if we alternative employ the partial-covering scenario.
Except for this \texttt{comptt},
the model composition is almost the same as that employed by \citet{Miniutti2007}. 
One of the  three gaussians expresses the narrow Fe-K emission line around 6.4 keV,
while the other two (with negative normalizations) the resonance absorption lines by ionized iron. 
The column density of neutral absorption was left free (see below).
We fixed the emissivity index of \texttt{laor} at 3.0 again.
Referring to the result obtained in section 3.4, 
the optical depth $\tau$ and the electron temperature of \texttt{comptt}
were fixed at 5.3 and 16.0 respectively,
while its normalization was left free.
The cutoff energy of \texttt{cutoffpl} was fixed at 200 keV, 
and its photon index $\Gamma$ was tied  to that of \texttt{pexrav}. 
The relative abundance of heavy elements other than Fe was fixed at 1.0,
while that of Fe, $A_{\rm Fe}$, was left free. 
The other parameters were left free.

With this model, we successfully obtained an acceptable fit ($\chi^2/$d.o.f. = $172.44/157$).  
The obtained parameters are summarized in table 1 (3rd column), 
and the best-fit model and residuals are shown in figure 8.
The column density, $N_{\rm H} \sim 0.8 \times 10^{22}$ cm$^{-2}$, 
is considered to approximate (in the limited anergy range of $> 2.5$ keV) the effect of warm absorber. 
Except for the reflection fraction $f_{\rm refl}$ and the inner radius $R_{\rm{in}}$, 
the derived parameters are consistent with those in \citet{Miniutti2007}. 
Since \texttt{comptt} accounts for some fraction of the hard X-ray  signals, 
the reflection intensity has decreased to $f_{\rm refl} \sim 1.2$, 
from the solution with $f_{\rm refl} > 2$ by  \citet{Miniutti2007}. 
The obtained inner radius of \texttt{laor} is $R_{\rm{in}} \sim10\;R_{\rm{g}}$, 
again in contrast to the value of  $R_{\rm{in}} \sim 2\;R_{\rm{g}}$ derived by \citet{Miniutti2007}. 
This is because the inclusion of the secondary hard component 
has slightly changed the continuum curvature around 3--6 keV.  
The equivalent width of the broad iron line is $ = 145^{+93}_{-74}$ eV, 
which is much smaller than $\sim 320$ eV in \citet{Miniutti2007}, 
but consistent with the value of $\sim 160$ eV expected 
from the reflection fraction of $f_{\rm refl} \sim 1.2$ (George \& Fabian 1991).

\subsection{Fitting with a relativistically burred reflection}


\begin{figure*}[t]
  \begin{center}
    \FigureFile(150mm,150mm)
     {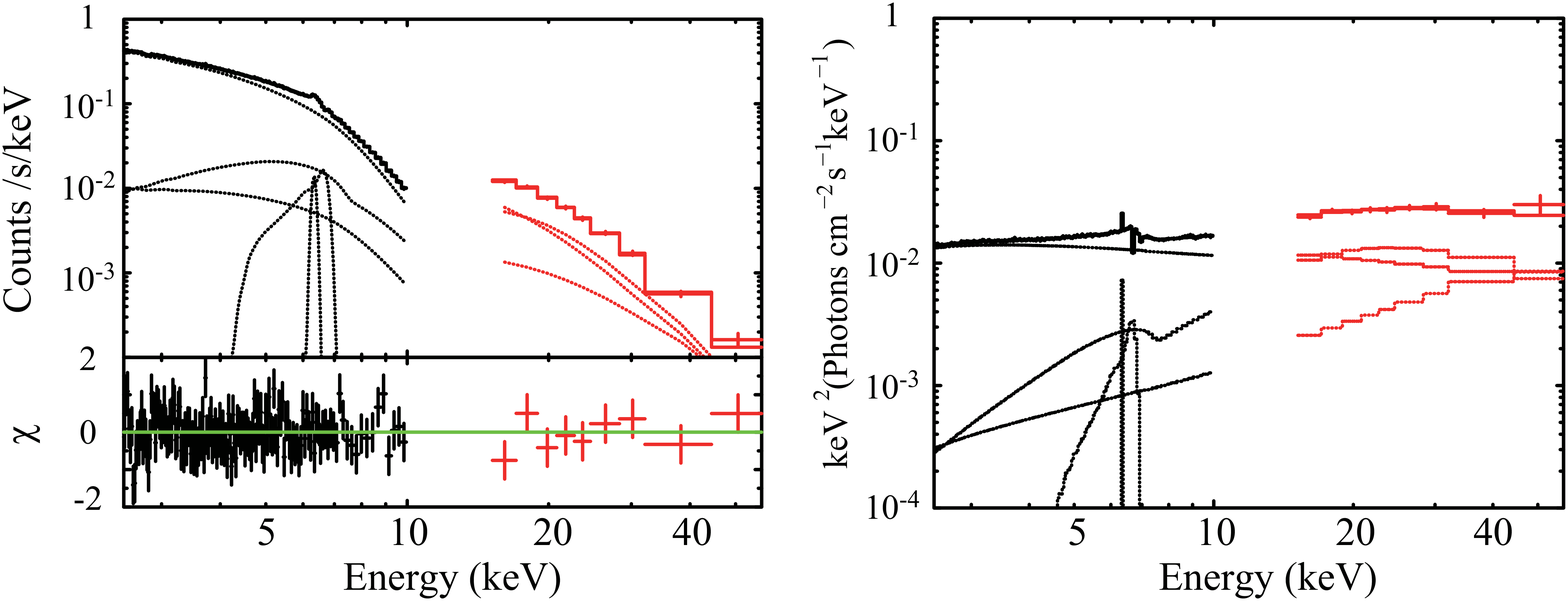}\\
  \end{center}
  \caption
 {The same as figure 6, but when the model is constructed as
 \texttt{wabs * (cutoffpl + kdblur * pexrav + comptt + laor + three narrow gaussians).}}
 \label{Test}
\end{figure*}


Although we succeeded in obtaining an acceptable model in section 3.1, 
it is physically still incomplete and self-inconsistent in the sense 
that the relativistic effects are applied only to the Fe-K line, 
but not to the reflection component which is expected to arise 
from almost the same locations as the Fe-K line.
Therefore, we further included the relativistic effect in \texttt{pexrav}. 
The model is now \texttt{wabs * (cutoffpl + kdblur * pexrav + comptt + laor + three narrow gaussians)}. 
Here, \texttt{kdblur}, implemented in xspec12, 
works as a convolution model  to blur a spectrum by reletivistic effects in  the same way as \texttt{laor}. 
The parameters of \texttt{kdblur} are all linked to those of \texttt{laor}. 
With the other parameters kept in the same condition as in section 4.1, 
we refitted the spectrum. 

As a result of this model improvement, we obtained a significantly better fit ($\chi^{2}/$d.o.f.  = $150.84/157$).
The result of this fitting is shown in figure 9, 
and the obtained parameters are shown in table 1 (4th column).
The fit parameters have remained, within errors, mostly the same as before.
In particular, the inner disk radius was again found at $R_{\rm in} \sim 10\;R_{\rm g}$ as before.
Although $f_\mathrm{refl}$ somewhat increased, it is still small enough to allow straightforward interpretation.
Thus, the spectrum can be interpreted invoking neither the extremely broad Fe-K line, 
nor the too strong reflection.
Since this solution does not require the accretion disk to intrude 
inside  the $6\;R_{\rm g}$ limit for a Schwarzschild BH, 
the Suzaku data of MCG--6-30-15 can be interpreted without invoking a rapidly spinning BH.

\section{Discussion and Summary}

Using the Suzaku XIS and HXD data acquired
on three occasions in 2006 January,
we studied broad-band X-ray spectra and variations of MCG--6-30-15.
On the first two occasions,
the hard X-ray  (15.0--45.0 keV) intensity variations
at 10 ks binning were consistent with being fully
correlated with those in the 3.0--10.0 keV energy band.
This means that the overall source variability
can be explained solely by intensity changes of the dominant power-law continuum in these periods.
The constant offsets in equation (1) and (2), 
or equivalently, the positive y-intercepts in figure 3,
are interpreted as the non-varying reflection component.
In the 3rd observation, in contrast,
the hard X-ray variations on the 10--50 ks time scales
were partially uncorrelated with those in the soft X-ray band.
In other words, the overall variability in this case
involved at least two independent parameters.

In the previous studies of MCG--6-30-15,
the hard X-ray spectral bump (entirely attributed to reflection)
was reported to remain rather constant
while the main power-law varied.
These results, however, were obtained mainly by
taking a difference spectrum between two time phases,
wherein the overall intensity is high and low
(e.g., figure 11 of Miniutti et al. 2007).
In our figure 2,
this means taking a difference between the
right and left halves of each panel.
By doing so, the newly discovered secondary variability
would approximately cancel out,
because of its independent nature,
and would not appear in the difference spectrum.
In that sense, the CCPs have played an essential role.

The source behavior in the 3rd observation has
been explained successfully by considering
that the spectrum involves a very hard spectral component
that is prominent in the HXD-PIN band and
varies independently of the power-law.
In the 3--45 keV energy range,
this ``variable hard component" could be approximated by
a power-law with $\Gamma \sim 1.0$.
More physically, this component can be interpreted 
in terms of a partial covering model.
We may then envisage that a thick absorber covers
a fraction of the entire power-law emission region.
Although such a strongly absorbed continuum must be accompanied 
by a strong fluorescent Fe-K line 
with an equivalent width of $\sim 2 \times (\alpha/4\pi)$ keV (Makishima et al. 1986),
where $\alpha$ is the solid angle of the absorption, 
the absorbed continuum itself carries only $\sim 1/30$ of the overall continuum at $\sim 7$ keV (figure 7).
Therefore, the associated Fe-K line is expected to have an equivalent width of 
$\sim 70 \times (\alpha/4\pi)$ eV.
Because the equivalent width of the narrow Gaussian was obtained as $23^{+5}_{-4}$ eV, we infer $\alpha/4\pi \sim 0.3$.
When the power-law intensity varies
with the covering fraction kept constant,
the source will move on the CCP along the correlation line.
When, instead, the covering fraction varies
under a constant power-law intensity,
only the soft X-ray intensity  will change,
and hence the source will move horizontally on the CCP.
Thus, in this interpretation, the covering fraction slightly changed
only in the 3rd observation while little in the first two.

An alternative idea of interest is to regard the variable hard emission
as a thermal Comptonization component
with an optical depth of $\tau \sim 5.3$
and an electron temperature of $T_{\rm e} \sim 16.0$ keV.
In this case, variations of this component is expected to cause
the source to move almost vertically on a CCP.
Although it is not obvious
whether such a relatively cool and dense Comptonizing plasma
can exist in an AGN,
an ingredient like this is not necessarily alien to the 
general view of accreting black holes.
Using the Suzaku data,
Makishima et al. (2008) in fact showed
that the Comptonizing corona in Cyg X-1 is highly inhomogeneous,
requiring at least two optical depths for Comptonization.
One possible candidate for a large value of $\tau$ is a kind
of transition zone between the disk and the surrounding corona.

Taking the newly identified hard X-ray variation into account,
the time-averaged 2.5--55 keV spectra of MCG--6-30-15
have been reproduced successfully using  7 components;
\texttt{comptt} (or equivalently, a partially-absorbed $\Gamma = 2.0$ power-law)
, a power-law with $\Gamma = 2.24$,
a reflection with $f_{\rm refl} \sim 1.7$,
a mildely boradened Fe-K line emitted from a disk
with $R_{\rm in} \sim  10\;R_{\rm g}$, and three Gaussians.
This result does not change even if
we apply relativistic smearing to the reflection component.
This condition implies a case
wherein the AGN in MCG--6-30-15 is a Schwarzschild black hole.

As originally pointed by Inoue and Matsumoto (2003) with ASCA, 
photo-ionized warm absorbers 
can reduce the need for the extreme Fe-K line broadening.
This has recently been demonstrated by Miller et al. (2008) and Miyakawa et al. (2009)
with the same Suzaku data as we used.
They have shown that the inclusion of warm absorbers 
can increase the inner disk radius derived from the Fe-K line shape.
It is our future task to more solidly combine the present results with the warm absorber scenario,
and hence to improve our understanding of the underlying continuum shape of this important AGN.

\appendix
\section{Principal Component Analysis}


\begin{table*}[t]
 \caption{Eigen values and eigen vectors of 1st to 3rd principal component.}
 \label{all_tbl}
 \begin{center}
  \begin{tabular}{cccc}
   \hline\hline
   Principal component & Eigen value & Eigen vector \\

   \hline
   
   1st   & 0.89 & $\left( \begin{array}{c}
   			0.46,~0.47,~0.47,~0.46,~0.37
			\end{array}
   			\right)$
   
   \\[1.5ex]

   2nd   & 0.09 & $\left( \begin{array}{c}
   			-0.28,
			-0.18,
			-0.17,
			-0.10,~0.92
			\end{array}
   			\right)$
   
   \\[1.5ex]

   3rd   & 0.01 & $\left( \begin{array}{c}
   			-0.51,
			-0.29,
			~0.08,
			~0.80,
			-0.11
			\end{array}
   			\right)$
   
   \\[1.5ex]
 \hline\hline
  \end{tabular}
 \end{center}
 \end{table*}



\begin{figure*}[t]
  \begin{center}
    \FigureFile(80mm,80mm)
     {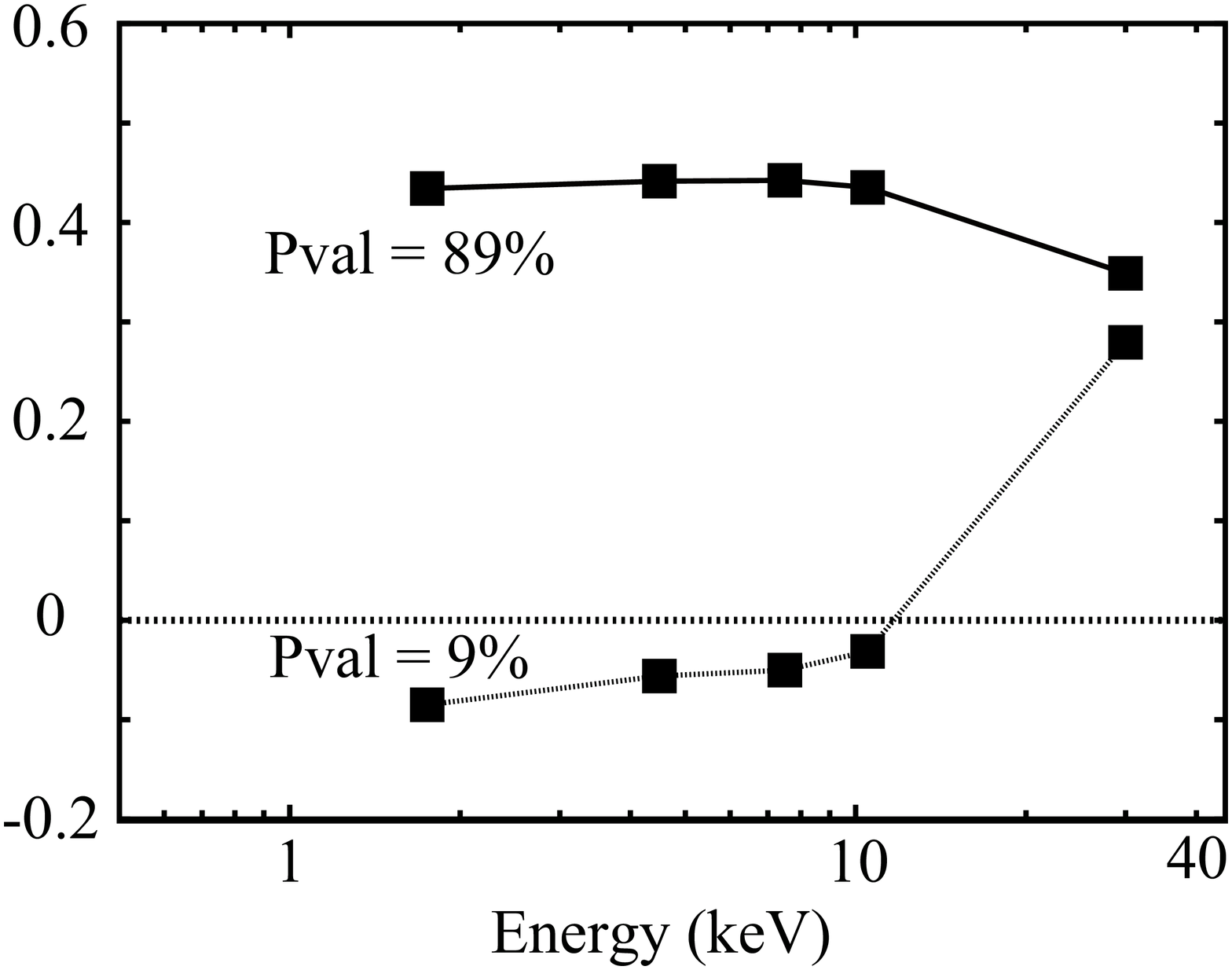}\\
  \end{center}
  \caption
 {The  first (upper data points) and the second (lower) eigenvectors
 from the principal component analysis, 
divided by the corresponding principal values.}
 \label{Test}
\end{figure*}


We carried out principal component analysis for the third interval.
The background-subtracted light curves, 
with a bin width of 10 ks,
were produced in 5 energy bands, 0.5--3.0 keV, 3.0--6.0 keV, 6.0--9.0 keV, 9.0--12.0 keV and 15.0--45.0 keV.
Each band has an error within 4\%.
At the $i$-th time bin, we can define a 5-demensional vector $\vec{v_{i}} = (C_{i1},~C_{i2},\cdots,~C_{i5})$,
where $C_{ij}$ is the counts at this time bin in the $j$-th energy band.
The ensemble $\{\vec{v_{i}}\}$ then forms a subset in a 5-dimensional Euclidean space.
The principle component analysis is a standard procedure to find principal axes (up to 5) of  $\{\vec{v_{i}}\}$,
and to estimate elongations (or ``principal values") of  $\{\vec{v_{i}}\}$ along those axes.

As a result of this analysis, 
the first and the second principal components were found to be dominant while the others are negligible.
Therefore, the ensemble  $\{\vec{v_{i}}\}$  has essentially a two-ddimensional distribution,
in agreement with the conclusion derived from figure 2c.
Table 2 shows eigen values and eigen vectors of the 1st to 3rd principal components. 
Figure 10 shows the  first and the second eigenvectors, 
divided by the corresponding principal values.
Thus, the first principal component has a relatively flat spectral distribution, 
and can be interpreted as variations of the most dominant power-law component.
In contrast, the 2nd principal component is concentrated in the highest energy range,
and interpreted as representing a very hard component that is ``orthogonal" to
(i.e., varying independently of) the 1st component. 
This is approximately identified with the ``variable hard component" discovered via CCP in section 3.3.


\begin{thebibliography}{}

\bibitem[Boldt (1987)]{Boldt1987}Boldt, E. 1987, Phys. Rep., 146, 215

\bibitem[Fukazawa \etal (2009)]Fukazawa Y., et al. 2009, PASJ, 61, S17

\bibitem[George \& Fabian (1991)]{GeorgeFabian1991}George, I. M., \&  Fabian, A. C. 1991, MNRAS,  249,  352

\bibitem[Gruber \etal(1999)]{Gruber1999}Gruber, D. E., Matteson, J. L., Peterson, L. E., \& Jung, G. V. 1999, ApJ, 520, 124

\bibitem{Inoue2003}Inoue, H., \& Matsumoto, C. 2003, PASJ, 55, 625-629

\bibitem[Ishisaki \etal (2007)]{Ishisaki2007}Ishisaki, Y., et. al. 2007, PASJ, 59, S113

\bibitem[Koyama \etal (2007)]{Koyama07}Koyama, K., et al. 2007, PASJ, 59, S23

\bibitem{Laor}Laor, A. 1991, ApJ, 90L, 376

\bibitem[Magdziarz \& Zdziarski (1995)]{MagdziarzZkziarski1995}Magdziarz, M., \& Zdziarski, A. 1995, MNRAS, 273, 837

\bibitem[Makishima \etal (2008)]{Makishima2008}Makishima, K., et al. 2008, PASJ 60, 585

\bibitem{Makishima1986}Makishima, K. 1986, in The Physics of Accretion onto Compact Objects, ed. K. O. Mason, M. G. Watson, \& N. E. White (Berlin: Springer) 249

\bibitem{Matsumoto2003}Matsumoto, C., Inoue, H., Fabian, A. C., \& Iwasawa, K. 2003, PASJ, 55, 613

\bibitem [McHardy \etal (2005)]{McHardy2005}McHardy, I. M., Gunn, K. F., Uttley, P., \& Goad, M. R. 2005, MNRAS, 359,1469

\bibitem{L_Miller08}Miller, L., Turner, T. J., \& Reeves, J. N. 2008, A\&A 483, 437

\bibitem [Miller \etal (2009)]{L_Miller09}Miller, L., Turner, T. J., \& Reeves, J. N. 2009, MNRAS, 399, L69

\bibitem[Miniutti \etal (2007)]{Miniutti2007} Miniutti, G., et al. 2007, PASJ, 59, S315

\bibitem [Miyakawa \etal (2009)]{Miyakawa2009} 
Miyakawa, T., Ebisawa, K., Terashima, Y., Tsuchihashi, F., Inoue, H., \& Zycki, P.,2009, PASJ, 61, 6, 1355

\bibitem [Nandra \etal (2007)]{Nandra2007}Nandra, K., O' Neill, P. M., George, I. M., \& Reeves, J. N.  2007, 
MNRAS, 382, 194

\bibitem [Nied{\'z}wiecki \& {\.Z}ycki (2008)]{NiedzwieckiZycki2008} Nied{\'z}wiecki, A. \& {\.Z}ycki, P. T. 2008, MNRAS, 386, 2, 759

\bibitem [Reynolds \& Nowak (2003)]{ReynoldNowak2003} Reynolds, C., \&  Nowak, M., Physics Reports, 377, 6, 389

\bibitem [Taylor \etal (2003)]{Taylor2003}Taylor, R. D., Uttley, P., \& McHardy, I. M., 2003, MARAS, 342, 2, 31

\bibitem[Takahashi \etal (2007)]{Takahashi07}Takahashi T., et al. 2007, PASJ, 59, S35

\bibitem [Tanaka \etal (1995)]{Tanaka1995}Tanaka, Y., et al. 1995, Nature, 375, 659 

\bibitem {}Uehara, Y. Master's thesis,  University of Tokyo, 2009

\bibitem [Vaughan \etal (2004)]{Vaughan2004} Vaughan, S., Fabian, A. C., Iwasawa, K., \& Turner, A. K., 2004, Nuclear Physics B Proceedings Supplements, 132, 244

\end{thebibliography}
\end{document}